\documentclass[aps, prl, 10pt, twocolumn, superscriptaddress, noshowpacs, preprintnumbers, floatfix]{revtex4-1}

\usepackage[colorlinks=true,breaklinks=true]{hyperref}
\hypersetup{allcolors=[rgb]{0.0 0.0 1.0},linkcolor=[rgb]{0.75 0.05 0.05}}
\usepackage[dvipsnames]{xcolor}
\usepackage{mathtools}
\usepackage{amsmath}
\usepackage{amsfonts}
\usepackage{amssymb}
\usepackage{bbold}
\usepackage{multirow}
\usepackage{bm}
\long\def\exclude#1{}
\usepackage{orcidlink}
\usepackage{esint}
\usepackage{microtype}

\newcommand{\bD}{{\bf D}}

\newcommand{\bp}{{\bf p}}

\newcommand{\br}{{\bf r}}
\newcommand{\bk}{{\bf k}}

\newcommand{\bK}{{\bf K}}

\newcommand{\be}{{\bf e}}

\newcommand{\bv}{{\bf v}}

\newcommand{\GF}{G_{\rm F}}

\begin{document}

\title{
Lepton-Number Crossings are Insufficient for Flavor Instabilities
}

\author{Damiano F.\ G.\ Fiorillo \orcidlink{0000-0003-4927-9850}}
\affiliation{Deutsches Elektronen-Synchrotron DESY,
Platanenallee 6, 15738 Zeuthen, Germany}

\author{Georg G.\ Raffelt
\orcidlink{0000-0002-0199-9560}}
\affiliation{Max-Planck-Institut f\"ur Physik, Boltzmannstr.~8, 85748 Garching, Germany}

\begin{abstract}
In dense neutrino environments, the mean field of flavor coherence can develop instabilities. A necessary condition is that the flavor lepton number changes sign as a function of energy and/or angle. Whether such a crossing is also sufficient has been a longstanding question. We construct an explicit counterexample: a spectral crossing without accompanying flavor instability, with an even number of crossings being key. This failure is physically understood as Cherenkov-like emission of flavor waves. If flipped-lepton-number neutrinos never dominate among those kinematically allowed to decay, the waves cannot grow.
\end{abstract}

\maketitle

{\bf\textit{Introduction.}}---Flavor evolution of dense neutrinos differs uniquely from conventional flavor oscillations. At high densities, neutrinos can exchange flavor with one another via the coherent weak field they generate, rather than through collisions, effectively forming a collisionless neutrino plasma. The resulting collective flavor conversions (CFCs)~\cite{Pantaleone:1992eq,Samuel:1993uw, Samuel:1995ri, Duan:2005cp, Duan:2006an, Tamborra:2020cul, Volpe:2023met, Johns:2025mlm,Raffelt:2025wty} can drastically influence compact transient sources such as core-collapse supernovae \cite{Nagakura:2023mhr, Ehring:2023lcd, Ehring:2023abs, Wang:2025nii} and neutron star mergers \cite{Wu:2017drk, Qiu:2025kgy}, affecting both their dynamics and associated nucleosynthesis.  Consistently accounting for CFCs is therefore a key microphysics requirement for understanding compact transients; a representative list of recent works includes Refs.~\cite{Bhattacharyya:2020jpj, Nagakura:2022kic,  Xiong:2023vcm, Shalgar:2022lvv, Cornelius:2023eop, Froustey:2024sgz, Abbar:2024ynh, Abbar:2024nhz, Richers:2024zit,  Fiorillo:2024qbl, Nagakura:2025brr, Liu:2025tnf, Padilla-Gay:2025tko}. 

Such a consistent description remains elusive. The main \hbox{obstacle} is the nonlinear, multi-scale evolution of CFCs, which drives interactions among flavor fluctuations across widely disparate spatial and temporal scales, down to centimeters and picoseconds. Numerical methods cannot fully capture this inherently nonlocal behavior and must instead rely on local approximations \cite{Bhattacharyya:2020jpj, Zaizen:2022cik, Zaizen:2023ihz, Xiong:2023vcm, Xiong:2024pue, Abbar:2024ynh, Fiorillo:2024qbl} of uncertain validity. While the relaxation of instabilities can be described in a quasi-linear framework~\cite{Fiorillo:2024qbl, Fiorillo:2025npi}, this approach has only been applied schematically. A fully nonlinear theoretical framework is still to be developed.

However, even before the dynamics turns nonlinear, a key question is: what triggers a flavor instability in the first place? One clue comes from the insight that an instability requires a lepton-number crossing, meaning a transition in the neutrino energy and angle distribution from the dominance of one flavor to another~\cite{Dasgupta:2021gfs,Johns:2024bob, Fiorillo:2024bzm}. In this description, antineutrinos are counted as neutrinos with negative energy and lepton number. The necessity of this crossing is easily understood from lepton-number conservation~\cite{Johns:2024bob, Fiorillo:2024bzm}, which inhibits flavor exchanges if all neutrino modes carry the same lepton number. This proof is equivalent to conservation of angular momentum that prevents the flipping of interacting spins if initially they all point in the same direction.

Whether such crossings are also sufficient remains an unsolved mystery. In the limit of massless neutrinos, the fast flavor regime, an angular crossing in the energy-integrated angular distribution guarantees an \hbox{instability}~\cite{Morinaga:2021vmc, Fiorillo:2024bzm}. While a spectral crossing (a lepton-number crossing in the energy distribution along some direction) does not trigger fast instabilities, it typically spawns slow ones, driven by mass-induced energy splitting. But is this always true? This question is gaining renewed urgency beyond purely mathematical interest, amid growing suspicion that slow modes have been systematically underestimated. They may, in fact, play a crucial role \hbox{\cite{Fiorillo:2024pns, Fiorillo:2025ank, Padilla-Gay:2025tko}}, particularly because they are the first instabilities to appear in compact transients~\cite{Fiorillo:2025gkw}.

A recent study of sufficiency \cite{Dasgupta:2025quc} has deduced certain conditions, but these are formal, depending on a modified distribution, which itself depends on the unknown eigenfrequency, and therefore is not conclusive. On the other hand, the conjecture of ergodicity suggests that the system is always unstable when it is not protected by lepton-number conservation~\cite{Johns:2024bob}.

The main goal of this Letter is to construct explicit counterexamples in the form of isotropic distributions with two spectral crossings or, more generally, with an even number of crossings. Multiple crossings can, in fact, occur in practice and were studied long ago in the context of homogeneous systems \cite{Dasgupta:2009mg, Raffelt:2011yb} (see also Ref.~\cite{Bhattacharyya:2021klg} for multiple angular crossings and their impact on fast instabilities).

In addition to identifying such crossed yet stable distributions, understanding their nature is an intriguing subject in its own right. We build on our recently developed picture of flavor instabilities as coherent emission of flavor waves, which are independent excitations whose quanta we have termed flavomons~\cite{Fiorillo:2024bzm, Fiorillo:2024uki, Fiorillo:2024pns, Fiorillo:2025ank, Fiorillo:2025npi, Fiorillo:2025zio}. In this framework, an instability corresponds to stimulated flavomon emission, and the instability condition arises purely from the breaking of detailed balance: neutrinos emit more waves than they absorb.

\textbf{\textit{Dispersion relation.}}---Flavor waves are neutrinos transitioning from one flavor to another. This collective motion, supported by the dense medium, has a frequency $\omega$ and a wave vector $\bk$, connected by a familiar dispersion relation. A derivation in the flavomon language is provided in the End Matter, whereas here we present only the final result for an axially symmetric medium, which serves as the starting point for our construction.

We denote the neutrino differential energy and angle distribution by $F_{\nu_\alpha}=dn_{\nu_\alpha}/dEdv$, with energy $E$,  flavor $\alpha$, and $v=\cos\theta$ direction relative to the symmetry axis. We take $\bk$ to be along the symmetry axis $\be_z$, so $\bk=k\be_z$ and so $k$ can be both positive and negative. What only matters is the difference in lepton number (DLN), which is defined, in a two-flavor system, as
\begin{equation}
    D(E,v)=\frac{1}{n_\nu}\begin{cases}
F_{\nu_e}(E, v) - F_{\nu_\mu}(E, v), &  E > 0; \\
F_{\overline{\nu}_\mu}(-E,v) - F_{\overline{\nu}_e}(-E, v), &  E < 0.
\end{cases}
\end{equation}
We have extracted the total number density $n_\nu$ to obtain a dimensionless function. Antineutrinos are counted as a species with negative energy and lepton number. 

The dispersion relation itself depends on integrals of the form
\begin{equation}\label{eq:integrals}
    I_n=\mu\int dv\,dE\, \frac{D(E,v)\,v^n}{\omega-kv-\omega_E+i\epsilon},
\end{equation}
where  $\omega_E=\delta m^2\cos 2\vartheta/2E$ is the energy splitting caused by the masses, the mixing angle is $\vartheta$, and the refractive scale is $\mu=\sqrt{2}\GF n_\nu$ with the Fermi constant $\GF$. For this study, the value of $\mu$ is inessential, but we mention that for typical supernova conditions, $n_\nu\sim 10^{33}\,\mathrm{cm}^{-3}$ so that $\mu\sim 10\,\mathrm{cm}^{-1}\sim 10^{-1}\,\mathrm{meV}$. We always use natural units with $\hbar=c=1$. The term $i\epsilon$ is a reminder of the retarded prescription to deal with solutions that have $\mathrm{Im}(\omega)\leq0$. Notice the different notation from Ref.~\cite{Fiorillo:2025zio}, where we denoted $D(E,v)=G(E,v)$ for $E>0$ and by $D(E,v)=\overline{G}(-E,v)$ for $E<0$. The present notation, which unifies positive and negative energies, is better suited here.

There are longitudinal and transverse flavomons. The former are axially symmetric modes that obey
\begin{equation}\label{eq:dispersion_longitudinal}
    (I_0-1)(I_2+1)-I_1^2=0,
\end{equation}
admitting two separate solutions. Transverse flavomons are axial-breaking modes and obey
\begin{equation}\label{eq:dispersion_transverse}
    I_0-I_2+2=0.
\end{equation}
Unstable modes are solutions of these dispersion relations with $\mathrm{Im}(\omega)>0$. For our simplest possible example we will actually assume an isotropic distribution, which is axisymmetric for any chosen $\bk$, and then all solutions are provided by Eqs.~\eqref{eq:dispersion_longitudinal} and~\eqref{eq:dispersion_transverse}.

\textbf{\textit{Explicit counterexample.}}---Our simple construction uses thermal energy distributions of the form
\begin{equation}
    F_{\nu_\alpha}(E)=\frac{\mathcal{N}_{\nu_\alpha}}{2T_{\nu_\alpha}}\left(\frac{E}{T_{\nu_\alpha}}\right)^2 e^{-E/T_{\nu_\alpha}}.
\end{equation}
We use nonvanishing populations only for $\nu_e$, $\nu_\mu$, and $\overline{\nu}_\mu$ with a rationale that will become clear later. The $\nu_e$ are taken to dominate, with $T_{\nu_e}=1\,\mathrm{MeV}$ and $\mathcal{N}_{\nu_e}=0.8$, with purely nominal units. The $\overline{\nu}_\mu$ and $\nu_\mu$ populations are subdominant, with $T_{\overline{\nu}_\mu}=T_{\nu_\mu}=3\,\mathrm{MeV}$, $\mathcal{N}_{\nu_\mu}=0.1$, and $\mathcal{N}_{\overline{\nu}_\mu}=0.05$. These choices are not meant to mimic realistic conditions, but to maximize the intended effect. For the unstable example, we reduce $\mathcal{N}_{\overline{\nu}_\mu}$ to 0.02. We express all results in units of $\mu$, so its actual value is inessential. The relative values of $\omega_E/\mu\sim 10^{-3}$--$10^{-2}$ are somewhat large compared with realistic values of $\omega_E/\mu\sim 10^{-5}$, but, as shown in Refs.~\cite{Fiorillo:2024pns,Fiorillo:2025ank,Fiorillo:2025zio}, the regime of instability is identical provided that $\omega_E\ll \mu \epsilon^2$, where $\epsilon$ is the ratio between the DLN and the total number of neutrinos, which in our case is $\epsilon \sim 1$. Hence, a larger $\omega_E/\mu$ simplifies the numerical solution without affecting the nature of the instability. We stress that all these inequalities and approximate equalities are to be interpreted as valid for the absolute value of the parameters: $\omega_E$ can be either positive or negative, but for the applicability of different approximations only its absolute magnitude in comparison with the other energy scales of the problem matters.

\begin{figure}
\includegraphics[width=1\columnwidth]{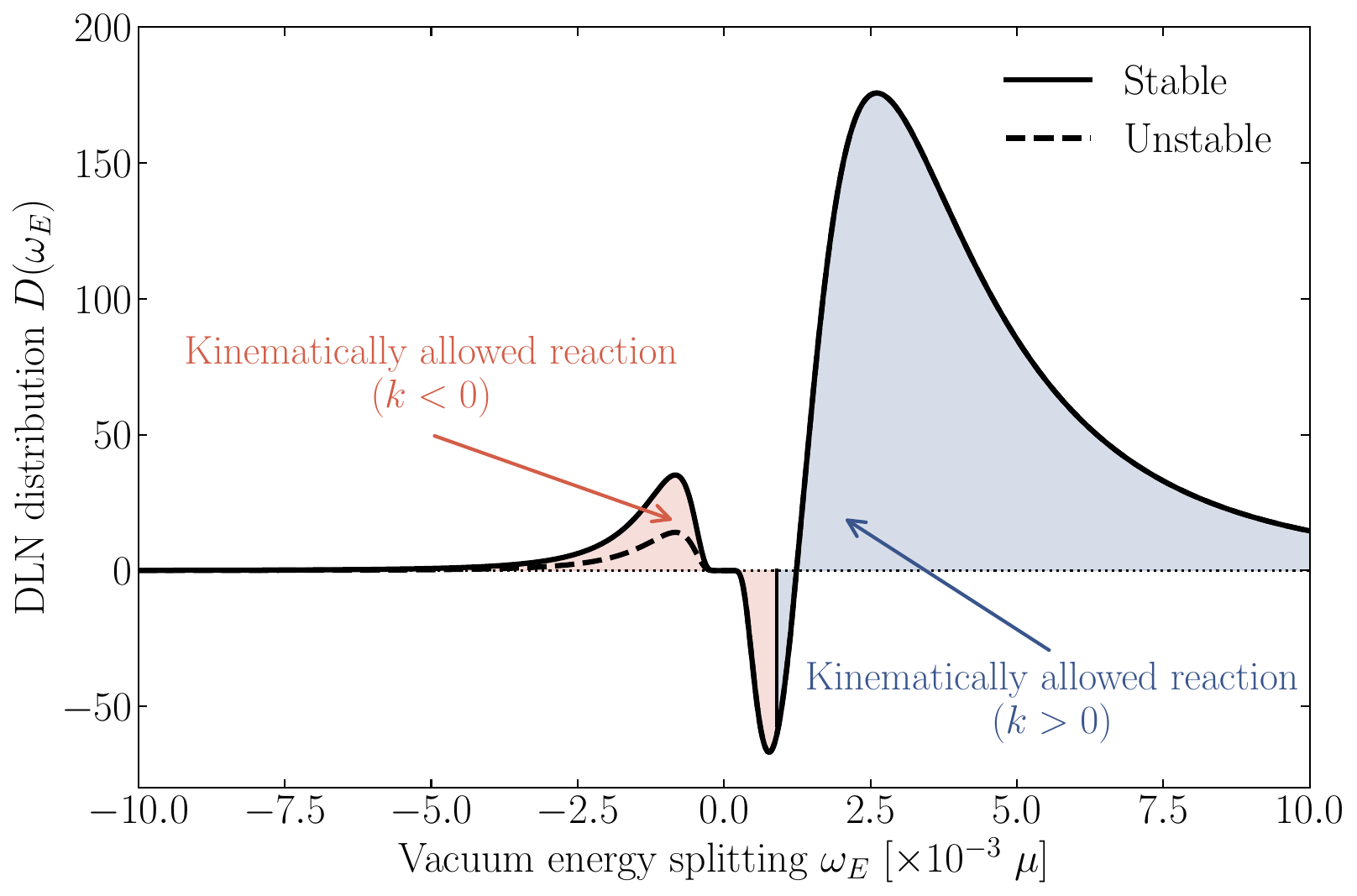}
\caption{DLN distributions for the stable (solid line) and unstable (dashed line) double-crossed setup. The vacuum energy splitting is $\omega_E=10^{-2}\, \mu\,(1\,\mathrm{MeV}/E)$. The DLN is shown in terms of $\omega_E$, so $D(\omega_E)=D(E)\,E/\omega_E$. For a given shifted frequency $\omega$ and shifted wave number $k$, all neutrinos below (above) the threshold $\omega_{E,\rm thr}=\omega-k$ can decay for $k<0$ ($k>0$).}\label{fig:stable_unstable}
\end{figure}

The stable and unstable DLN distributions are shown in Fig.~\ref{fig:stable_unstable}. Both are double crossed. At negative $\omega_E$, they are positive, due to $\overline{\nu}_\mu$ carrying negative muon lepton number. At large positive $\omega_E$ (small neutrino energies), the distributions are also positive, due to the dominance of $\nu_e$. However, at small positive $\omega_E$ (high neutrino energies), they are negative, due to the larger $T_{\nu_\mu}$. 

\begin{figure}
    \centering
    \includegraphics[width=\columnwidth]{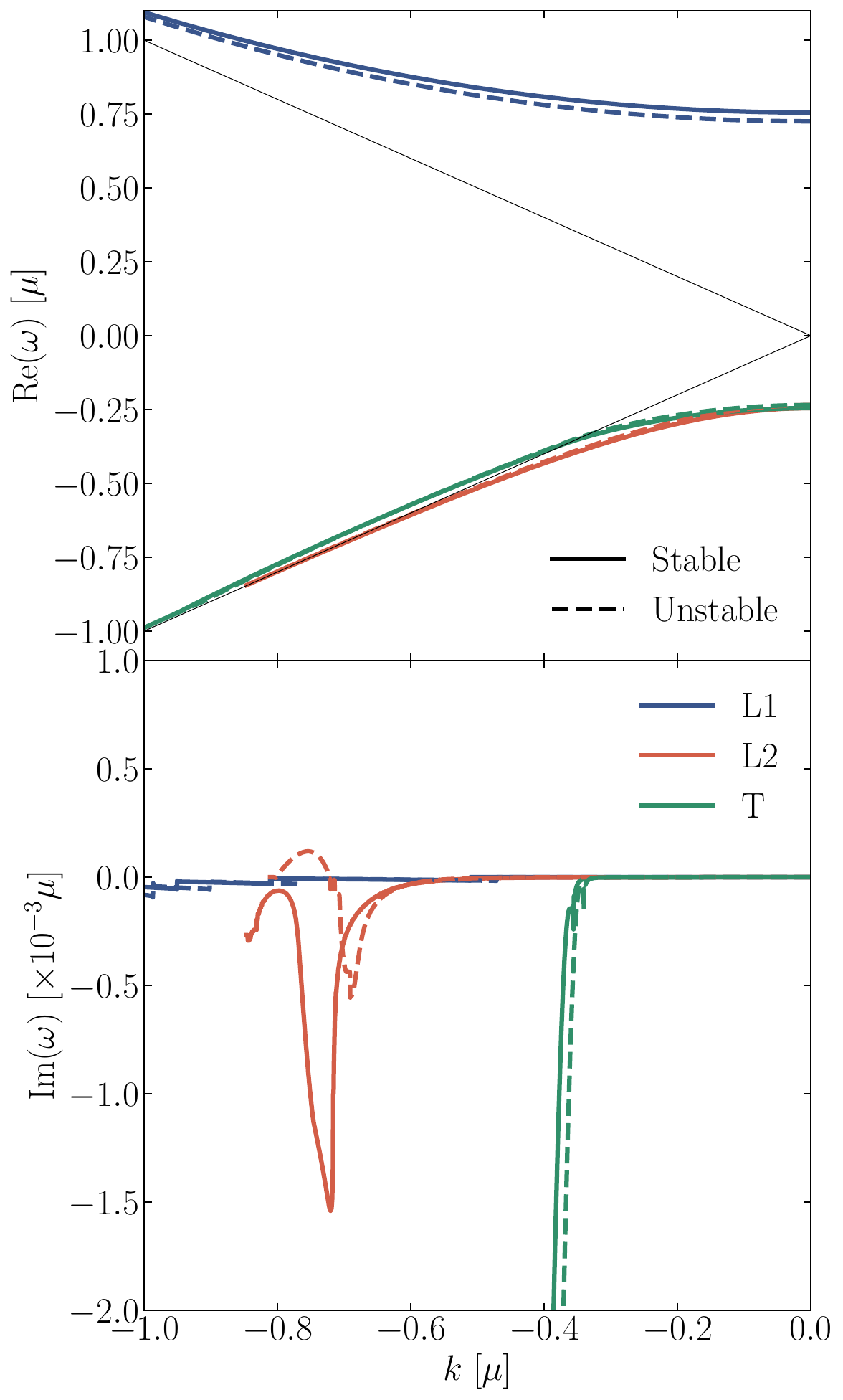}
    \caption{Dispersion relations for the stable (solid) and unstable (dashed) distributions. Different colors for the different solutions, namely the two longitudinal modes (L1 and L2) from Eq.~\eqref{eq:dispersion_longitudinal} and the transverse mode (T) from Eq.~\eqref{eq:dispersion_transverse}. The light cone $\mathrm{Re}(\omega)=\pm k$ is shown as thin black lines.}
    \label{fig:modes}
\end{figure}

We next seek explicit numerical solutions of the dispersion relations in Eqs.~\eqref{eq:dispersion_longitudinal} and~\eqref{eq:dispersion_transverse}. Following our previous technique \cite{Fiorillo:2024pns, Fiorillo:2024dik, Fiorillo:2025ank, Fiorillo:2025zio}, we use a root-finding algorithm on a finely spaced grid in $k$, using a given solution as seed for the next search. We begin at $k=0$, where the gapped superluminal solution is analytically known \cite{Fiorillo:2025zio}. We do not search for gapless modes  which are generally Landau damped and there is no analytical guidance about their existence or properties \cite{Fiorillo:2025zio}. A purely numerical search did not return any such solution. In any case, our later arguments explain theoretically the absence of an instability, so this numerical search is mostly illustrative.

Figure~\ref{fig:modes} shows the solutions in term of the real and imaginary parts of their frequency as a function of $k$, where L1 and L2 are the two longitudinal branches, whereas T is the transverse one. Isotropy ensures that $\omega(\bk)$ is the same for equal $|\bk|$ and we have resolved to show the solution for $k<0$ as for $k>0$ the waves move ``backward,'' which is slightly less intuitive. The real part of $\omega$ shows the familiar structure \cite{Fiorillo:2024pns, Fiorillo:2025ank, Fiorillo:2025zio}. All branches begin at a finite $\mathrm{Re}(\omega)$ at $k=0$ (hence the name gapped~\cite{Fiorillo:2025zio}), and rapidly approach the light cone, corresponding to flavor waves moving with the speed of light. For the L2 and the T mode, this is the positive light cone, i.e., waves moving forward, while for the L1 mode the waves move backward. Crucially, the branches also cross the light cone, becoming kinematically able to interact with neutrinos with negative $\omega_E$.

More important is the imaginary part, the growth or damping rate. Using the retarded dispersion relation, which includes the $i\epsilon$ prescription, allows us to find solutions with a negative $\mathrm{Im}(\omega)$, Landau-damped flavor modes, that we first identified in Refs.~\cite{Fiorillo:2023mze, Fiorillo:2023hlk} and later explained, in general, in Ref.~\cite{Fiorillo:2024bzm}. The often used version without $i\epsilon$ can only reveal growing modes and superluminal real-valued ones. Crucially for our present discussion, there is a continuity of solutions from negative to positive $\mathrm{Im}(\omega)$ when distorting the DLN distribution.

In any case, Fig.~\ref{fig:modes} reveals that for the stable case, all modes remain Landau damped for all $k$. The L2 branch exhibits a negative $\mathrm{Im}(\omega)$ that first drops and then increases, getting close to 0, but never turns positive. In contrast, our unstable example, only slightly distorted from the stable one, shows exactly that: the distortion is sufficient to push $\mathrm{Im}(\omega)$ above 0, creating a small unstable bump as function of $k$. These unstable modes would also show up without $i\epsilon$, but the branch would look like an isolated unstable bump at large $k$, difficult to find numerically. In this sense, the $i\epsilon$ prescription is of practical numerical relevance, because the branches are continuously connected to the $k=0$ solution.

The imaginary part of the branches abruptly terminates at a finite $k$, due to unavoidable singularities in the dispersion relation when the wave begins to move in phase with neutrinos of very low energy. Numerically, this manifests as a failure of the root-finding algorithm to satisfy the dispersion relation at larger $k$. 
We now turn to an analytical interpretation that captures the origin of this feature from first principles.

{\textbf{\textit{Flavomon picture of CFCs.}}}---The physical picture of instability relies on collective flavor excitations, which are classical flavor waves with quanta that are quasi-particles termed flavomons~$\psi$. An instability corresponds to stimulated $\psi$ emission through $\overline{\nu}_e\to \overline{\nu}_\mu+\psi$ and $\nu_\mu\to \nu_e+\psi$, where we assume $\nu_e$ dominance in the neutrino plasma, so the main flavomon excitations carry one unit of muon and minus 
one unit of electron lepton number. These decays are constrained by kinematics; the flavomon energy and momentum are given by their frequency $\Omega$ and wave vector $\bK=K\mathbf{e}_z$ along the symmetry axis, and we will now see that these are simply related to $\omega$ and $\bk$ introduced earlier. 

For our arguments, a crucial point is that a flavomon is \textit{not} accompanied by a corresponding antiflavomon state~\cite{Fiorillo:2025npi}. The symmetry between the two states is broken by the plasma having a preferred flavor lepton number. In particular, in a plasma containing only $\nu_e$ (and potentially $\overline{\nu}_\mu$), flavomon states can only carry the flavor lepton number of a $\overline{\nu}_e\nu_\mu$ pair, and antiflavomon states do not exist. This asymmetry was proven explicitly from the dispersion relation~\cite{Fiorillo:2025npi}, but can be understood physically. In a plasma dominated by $\nu_e$ along all directions, in the massless regime there cannot be any antiflavomon state. This is because the reactions $\nu_e\to \nu_\mu+\overline{\psi}$ and its inverse and $\overline{\nu}_\mu\to\overline{\nu}_e+\overline{\psi}$ and its inverse, all have the same matrix element and energy conservation conditions, and so the dominance of $\nu_e$ would cause emission to dominate over absorption. The resulting instability is forbidden by lepton-number conservation~\cite{Johns:2024esf, Fiorillo:2024bzm}. When neutrino masses are taken into account, the energy conservation conditions for the different reactions change, but the global properties of the flavomon states remain unchanged~\cite{Fiorillo:2025npi}. Hence, the flavomon energy levels are only mildly affected---their real energy and quantum number are the same---so only their growth rate changes due to the emission and absorption processes. 

The $\overline\nu_e$ or $\nu_\mu$ energy is split from that of the corresponding $\nu_e$ or $\overline\nu_\mu$ by neutrino refraction and the mass splitting. The energy splitting relevant for decay is $\Delta E=E_{\nu_e}-E_{\nu_\mu}=\mu (D_0-v D_1)-\omega_E$, where $D_n=\int dv dE\, D(E,v) v^n$ are the moments of the DLN. Therefore, in the decay $\nu_\mu\to \nu_e+\psi$, with an initial neutrino momentum~$\bp$, energy conservation implies~\cite{Fiorillo:2025npi}
\begin{equation}
    |\bp|=|\bp-\bK|+\Delta E+\mathrm{Re}(\Omega).
\end{equation}
Expanding for $\mathrm{Re}(\Omega),|\bK|\ll |\bp|$ yields
\begin{equation}\label{eq:kinematic_condition}
    \mathrm{Re}(\omega)-kv-\omega_E=0,
\end{equation}
where $\omega=\Omega+\mu D_0$ and $\bk=\bK+\mu D_1 \mathbf{e}_z$. In other words, Eq.~\eqref{eq:kinematic_condition} implies a vanishing real part of the denominator in Eq.~\eqref{eq:integrals} and only neutrinos satisfying 
Eq.~\eqref{eq:kinematic_condition} can decay.

For $\omega\gtrsim k$, neutrinos can never satisfy this condition because $\omega_E\ll \mu$. Neutrinos can decay only to near-luminal flavomons ($\omega\sim k$), corresponding to the instability appearing only for flavor waves close to the light cone in Fig.~\ref{fig:modes}. Neutrinos that can decay are then only those~with
\begin{equation}\label{eq:angle}
    v=\frac{\omega-\omega_E}{k},
\end{equation}
meaning for $k>0$ only those with $\omega_E>\omega-k$ and for $k<0$ with $\omega_E<\omega-k$. The neutrino distribution is therefore split into two sectors: one that can decay for $k<0$ (highlighted in red in Fig.~\ref{fig:stable_unstable}), and the other that can decay for $k>0$ (highlighted in blue).

\begin{figure}
    \centering
    \includegraphics[width=\columnwidth]{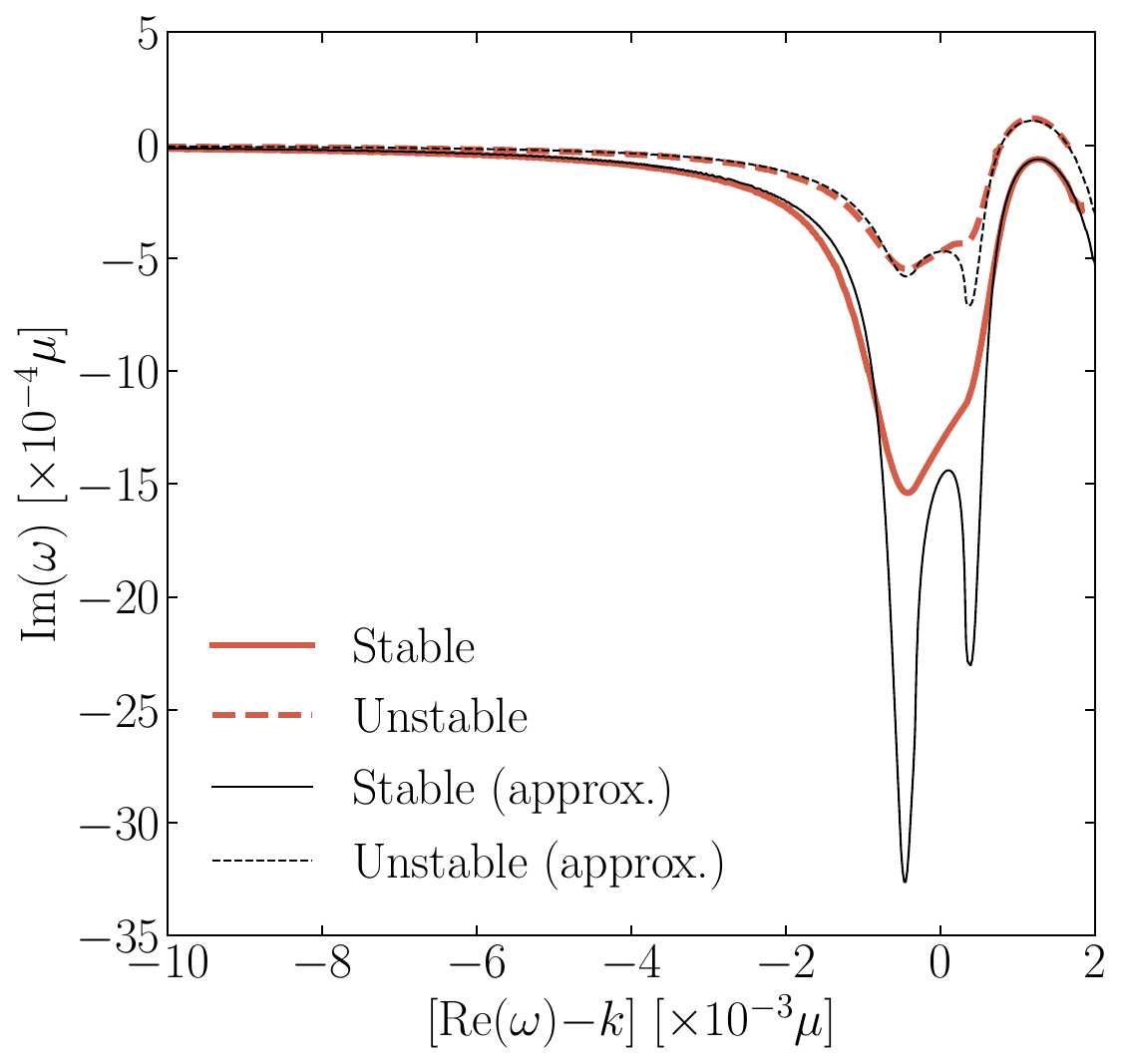}
    \caption{Comparison of the numerical growth rates with the analytical approximations, valid for $|\mathrm{Im}(\omega)|\ll |\mathrm{Re}(\omega)-k|$.}
    \label{fig:analytic}
\end{figure}

The crucial observation is then that the production rate of a given flavomon mode depends on the total neutrino population in the red or the blue region. For the example of Fig.~\ref{fig:stable_unstable}, neutrinos with positive DLN absorb flavomons via $\psi+\nu_e\to \nu_\mu$ and $\psi+\overline{\nu}_\mu\to \overline{\nu}_e$. Conversely, neutrinos with negative DLN produce flavomons via $\overline{\nu}_e\to \overline{\nu}_\mu+\psi$ or $\nu_\mu\to \nu_e+\psi$. For any given value of $\omega-k$, \textit{all} neutrinos with $\omega_E<\omega-k$ participate in flavomon production or absorption. The mode will only be unstable (for $k<0$) if the former dominate, i.e., if the red shaded area is overall negative, dominated by negative-DLN neutrinos. If for all values of $\omega-k$ it remains positive, all modes are damped, and the distribution is stable, despite its crossings.

This picture is directly confirmed by mathematics. The growth rate of a longitudinally polarized mode can be written explicitly as
\begin{equation}\label{eq:approx_growth}
    \mathrm{Im}(\omega)=\pi\,\frac{\int dE\,D(E,1) \Theta\bigl[\mathrm{Re}(\omega)-k-\omega_E\bigr]}{\fint dE \frac{D(E,1)}{\mathrm{Re}(\omega)-k-\omega_E}}
\end{equation}
for $k<0$. Here $\fint$ denotes the principal value of the integral. This approximate expression, first derived in Ref.~\cite{Fiorillo:2025zio}, is valid for very weak instabilities, and shows the natural appearance of the integrated DLN for all neutrinos satisfying the kinematic threshold. The denominator, which depends on a global integral of the neutrino energy distribution, is only mildly affected by the small fraction of flipped neutrinos for a weakly unstable distribution, and therefore its sign is fixed. Hence, an instability ensues if the integral in the numerator changes sign for some values of $k$. We review the derivation of this equation in the End Matter.

Figure~\ref{fig:analytic} compares the numerical growth rate for the L2 branch, the one eventually turning unstable, with the analytical result of Eq.~\eqref{eq:approx_growth}. The match is nearly perfect where the approximation is supposed to hold: for small growth rates. It also affirms our interpretation that the stable case is not protected by a global conservation law, but by kinematics. For any flavomon mode, neutrinos fulfilling the decay kinematics are outnumbered by the $\overline{\nu}_\mu$, corresponding to negative energies. Flavomons are produced by high-energy $\nu_\mu$, but absorbed in larger quantities by $\overline{\nu}_\mu$, causing overall stability. Reducing the $\overline{\nu}_\mu$ abundance, as we do to make our distribution unstable, suppresses flavomon absorption enough to cause~instability.

{\bf\textit{Discussion.}}---Understanding what triggers flavor instabilities is central to assessing the impact of collective flavor conversions. For years, attention has focused on fast instabilities, which occur even for massless neutrinos and are associated with angular crossings. However, it has become clear that these are not the first to appear in SNe; spectral crossings occur earlier~\cite{Fiorillo:2025gkw}, meaning that slow instabilities probably set the stage. We have provided the first explicit example of a distribution with spectral lepton-number crossings that exhibits no instability, falsifying the idea that such crossings are always associated with unstable modes.

The construction of our counterexample was guided by the \textit{flavomon} picture of flavor instabilities. In this framework, an instability arises from an imbalance between neutrinos capable of emitting a given flavomon mode and those absorbing it: if emitters dominate, the mode grows. The kinematics of the neutrino-flavomon process reveals that only neutrinos with vacuum energy splitting above or below a critical threshold, $\omega_E = \mathrm{Re}(\omega) - k$, can participate. This criterion enabled us to design a distribution with spectral crossings, but no instability.

This understanding hinges on adopting the analytic dispersion relation, including the $i\epsilon$ prescription, that gives rise to Landau-damped modes with $\mathrm{Im}(\omega) < 0$. Otherwise one discards the quasi-particle interpretation of flavor waves and the instability mechanism cannot be understood as emission and absorption, and our result would lack a physical explanation.  The flavomon picture is central to uncovering the true dynamics underlying collective flavor conversions.

Multiple crossings are key to our counterexample. A single-crossed DLN distribution changes sign between $\omega_E\to-\infty$ and $\omega_E\to+\infty$, so one can always choose $\mathrm{Re}(\omega)-k$ such that the neutrinos contributing to flavomon emission have only negative DLN. The same holds for any distribution with an odd number of crossings, which seem to guarantee an instability. This result directly matters in SNe, where a $\overline{\nu}$ crossing, caused by an excess of $\overline{\nu}_e$ over $\overline{\nu}_\mu$, ensures an instability. On the other hand, a spectral crossing caused by an excess of high-energy $\nu_\mu$ over $\nu_e$, while the antineutrinos are dominated by $\overline{\nu}_\mu$, need not spawn an instability, as we have seen. 

Our reasoning yields a quantitative criterion for the instability and, even better, a direct expression for the growth rate (Eq.~\eqref{eq:approx_growth}). Its generality descends primarily from the observation that, close to an instability, the growth rate must pass through 0 (from Landau damped to unstable), and so must be very small. This is precisely the regime where Eq.~\eqref{eq:approx_growth} applies, so its implications are particularly robust in this respect. 

Our proof of nonexistence of unstable modes, besides the negative numerical search for the specific example, relies heavily on Eq.~\eqref{eq:approx_growth}, which holds only for gapped modes~\cite{Fiorillo:2025zio}, defined by $\omega,\, k\sim \mu \epsilon\gg \omega_E$; here $\epsilon$ is the flavor lepton-number asymmetry, defined as the total DLN of the plasma divided by the total number of neutrinos, not to be confused with the Landau $i\epsilon$ prescription. Putative gapless unstable modes with $\omega, \, k\sim \omega_E$ (numerically not found here) would be irrelevant, in practice, because of their wavelength exceeding the pertinent astrophysical length scales. In this sense, we have not delivered a mathematical derivation, but a physical argument applicable to the cases of practical~interest.

It is critical that we have used continuous, rather than discrete, energy and angle distributions. Discretized models, especially involving few beams, can introduce unphysical artifacts. For instance, a monodirectional neutrino beam lacks collective modes, while even a narrow spread restores them, e.g., in SNe, from  Eq.~\eqref{eq:angle}, a spread of $\theta \sim \sqrt{\omega_E/\mu\epsilon} \sim 10^{-3}$--$10^{-2}$. Likewise, spurious instabilities may arise from monochromatic energies combined with discrete angle bins~\cite{Sarikas:2012ad}. While one can construct apparently stable ``crossed'' discrete distributions, it is unclear if the stability is physical. Continuous distributions are thus essential to capture the onset of slow instabilities. This feature is known for plasma instabilities~\cite{boyd2003physics}, which at their onset depend on the continuity of the distribution because they are kinetic rather than hydrodynamical.

Crossed distributions without instability also contradict the intriguing idea of ergodicity~\cite{Johns:2024bob}, which holds that CFCs would arise when they are not prevented by lepton-number conservation, i.e., as soon as there is a crossing. Our counterexample points instead to a more dynamical picture of instability, not ergodic but kinetic, driven by a balance between the emission and absorption of flavomons.

{\bf\textit{Acknowledgments.}}---D.F.G.F.\ is supported by the Ale\-xander von Humboldt Foundation (Germany). G.G.R.\ acknowledges partial support by the German Research Foundation (DFG) through the Collaborative Research Centre ``Neutrinos and Dark Matter in Astro- and Particle Physics (NDM),'' Grant SFB--1258--283604770, and under Germany’s Excellence Strategy through the Cluster of Excellence ORIGINS EXC--2094--390783311.

\bibliographystyle{bibi}
\bibliography{References}

@article{Dasgupta:2009mg,
    author = "Dasgupta, Basudeb and Dighe, Amol and Raffelt, Georg G. and Smirnov, Alexei Yu.",
    title = "{Multiple Spectral Splits of Supernova Neutrinos}",
    eprint = "0904.3542",
    archivePrefix = "arXiv",
    primaryClass = "hep-ph",
    reportNumber = "MPP-2009-33",
    doi = "10.1103/PhysRevLett.103.051105",
    journal = "Phys. Rev. Lett.",
    volume = "103",
    pages = "051105",
    year = "2009"
}

@article{Padilla-Gay:2025tko,
    author = "Padilla-Gay, Ian and Chen, Heng-Hao and Abbar, Sajad and Wu, Meng-Ru and Xiong, Zewei",
    title = "{Flavor equilibration of supernova neutrinos: Exploring the dynamics of slow modes}",
    eprint = "2505.11588",
    archivePrefix = "arXiv",
    primaryClass = "astro-ph.HE",
    reportNumber = "SLAC-PUB-250421",
    doi = "10.1103/jg14-8p4l",
    journal = "Phys. Rev. D",
    volume = "112",
    number = "4",
    pages = "043039",
    year = "2025"
}

@article{Nagakura:2025brr,
    author = "Nagakura, Hiroki and Zaizen, Masamichi and Liu, Jiabao and Johns, Lucas",
    title = "{Resolution requirements for numerical modeling of neutrino quantum kinetics}",
    eprint = "2501.14145",
    archivePrefix = "arXiv",
    primaryClass = "astro-ph.HE",
    reportNumber = "LA-UR-24-33051",
    doi = "10.1103/PhysRevD.111.043028",
    journal = "Phys. Rev. D",
    volume = "111",
    number = "4",
    pages = "043028",
    year = "2025"
}

@article{Liu:2025tnf,
    author = "Liu, Jiabao and Nagakura, Hiroki and Zaizen, Masamichi and Johns, Lucas and Yamada, Shoichi",
    title = "{Asymptotic states of fast neutrino-flavor conversions in the three-flavor framework}",
    eprint = "2503.18145",
    archivePrefix = "arXiv",
    primaryClass = "astro-ph.HE",
    doi = "10.1103/v9lr-ydbb",
    journal = "Phys. Rev. D",
    volume = "111",
    number = "12",
    pages = "123004",
    year = "2025"
}

@article{Qiu:2025kgy,
    author = "Qiu, Yi and Radice, David and Richers, Sherwood and Bhattacharyya, Maitraya",
    title = "{Neutrino Flavor Transformation in Neutron Star Mergers}",
    eprint = "2503.11758",
    archivePrefix = "arXiv",
    primaryClass = "astro-ph.HE",
    reportNumber = "INT-PUB-25-004",
    doi = "10.1103/h2q7-kn3v",
    journal = "Phys. Rev. Lett.",
    volume = "135",
    number = "9",
    pages = "091401",
    year = "2025"
}

@article{Abbar:2024nhz,
    author = "Abbar, Sajad and Volpe, Maria Cristina",
    title = "{Using Bayesian inference to distinguish neutrino flavor conversion scenarios via a prospective supernova neutrino signal}",
    eprint = "2401.10851",
    archivePrefix = "arXiv",
    primaryClass = "astro-ph.HE",
    reportNumber = "MPP-2024-11",
    doi = "10.1103/PhysRevD.111.083005",
    journal = "Phys. Rev. D",
    volume = "111",
    number = "8",
    pages = "083005",
    year = "2025"
}

@article{Ehring:2023lcd,
    author = "Ehring, Jakob and Abbar, Sajad and Janka, Hans-Thomas and Raffelt, Georg and Tamborra, Irene",
    title = "{Fast neutrino flavor conversion in core-collapse supernovae: A parametric study in 1D models}",
    eprint = "2301.11938",
    archivePrefix = "arXiv",
    primaryClass = "astro-ph.HE",
    doi = "10.1103/PhysRevD.107.103034",
    journal = "Phys. Rev. D",
    volume = "107",
    number = "10",
    pages = "103034",
    year = "2023"
}

@article{Nagakura:2023mhr,
    author = "Nagakura, Hiroki",
    title = "{Roles of Fast Neutrino-Flavor Conversion on the Neutrino-Heating Mechanism of Core-Collapse Supernova}",
    eprint = "2301.10785",
    archivePrefix = "arXiv",
    primaryClass = "astro-ph.HE",
    doi = "10.1103/PhysRevLett.130.211401",
    journal = "Phys. Rev. Lett.",
    volume = "130",
    number = "21",
    pages = "211401",
    year = "2023"
}

@article{Ehring:2023abs,
    author = "Ehring, Jakob and Abbar, Sajad and Janka, Hans-Thomas and Raffelt, Georg and Tamborra, Irene",
    title = "{Fast Neutrino Flavor Conversions Can Help and Hinder Neutrino-Driven Explosions}",
    eprint = "2305.11207",
    archivePrefix = "arXiv",
    primaryClass = "astro-ph.HE",
    doi = "10.1103/PhysRevLett.131.061401",
    journal = "Phys. Rev. Lett.",
    volume = "131",
    number = "6",
    pages = "061401",
    year = "2023"
}

@article{Wang:2025nii,
    author = "Wang, Tianshu and Burrows, Adam",
    title = "{The Effect of the Fast-Flavor Instability on Core-Collapse Supernova Models}",
    eprint = "2503.04896",
    archivePrefix = "arXiv",
    primaryClass = "astro-ph.HE",
    month = "3",
    year = "2025",
       volume = {986},
       number = {2},
          eid = {153},
        pages = {153},
          doi = {10.3847/1538-4357/add889},
          journal={Astrophys. J.}
}

@article{Fiorillo:2024pns,
    author = "Fiorillo, Damiano F. G. and Raffelt, Georg G.",
    title = "{Theory of neutrino slow flavor evolution. Part I. Homogeneous medium}",
    eprint = "2412.02747",
    archivePrefix = "arXiv",
    primaryClass = "hep-ph",
    doi = "10.1007/JHEP04(2025)146",
    journal = "JHEP",
    volume = "04",
    pages = "146",
    year = "2025"
}

@article{Johns:2025mlm,
    author = "Johns, Lucas and Richers, Sherwood and Wu, Meng-Ru",
    title = "{Neutrino Oscillations in Core-Collapse Supernovae and Neutron Star Mergers}",
    eprint = "2503.05959",
    archivePrefix = "arXiv",
    primaryClass = "astro-ph.HE",
    reportNumber = "LA-UR-25-21809",
    doi = "10.1146/annurev-nucl-121423-100853",
    month = "3",
    year = "2025",
    journal = "Annu. Rev. Nucl. Part. Sci.",
    volume = "75"
}

@article{Froustey:2024sgz,
    author = "Froustey, Julien and Kneller, James P. and McLaughlin, Gail C.",
    title = "{Quantum maximum entropy closure for small flavor coherence}",
    eprint = "2409.05807",
    archivePrefix = "arXiv",
    primaryClass = "hep-ph",
    reportNumber = "N3AS-24-031",
    doi = "10.1103/PhysRevD.111.063022",
    journal = "Phys. Rev. D",
    volume = "111",
    number = "6",
    pages = "063022",
    year = "2025"
}

@article{Richers:2024zit,
    author = "Richers, Sherwood and Froustey, Julien and Ghosh, Somdutta and Foucart, Francois and Gomez, Javier",
    title = "{Asymptotic-state prediction for fast flavor transformation in neutron star mergers}",
    eprint = "2409.04405",
    archivePrefix = "arXiv",
    primaryClass = "astro-ph.HE",
    reportNumber = "N3AS-24-030",
    doi = "10.1103/PhysRevD.110.103019",
    journal = "Phys. Rev. D",
    volume = "110",
    number = "10",
    pages = "103019",
    year = "2024"
}

@article{Abbar:2024ynh,
    author = "Abbar, Sajad and Wu, Meng-Ru and Xiong, Zewei",
    title = "{Application of neural networks for the reconstruction of supernova neutrino energy spectra following fast neutrino flavor conversions}",
    eprint = "2401.17424",
    archivePrefix = "arXiv",
    primaryClass = "astro-ph.HE",
    reportNumber = "MPP-2024-12",
    doi = "10.1103/PhysRevD.109.083019",
    journal = "Phys. Rev. D",
    volume = "109",
    number = "8",
    pages = "083019",
    year = "2024"
}

@article{Fiorillo:2025npi,
    author = "Fiorillo, Damiano F. G. and Raffelt, Georg G.",
    title = "{Collective Flavor Conversions Are Interactions of Neutrinos with Quantized Flavor Waves}",
    eprint = "2502.06935",
    archivePrefix = "arXiv",
    primaryClass = "hep-ph",
    month = "2",
    year = "2025",
    journal = {Phys. Rev. Lett.},
    pages = "211003",
    volume = "134"
}

@article{Fiorillo:2024dik,
    author = "Fiorillo, Damiano F. G. and Goimil-Garc\'\i{}a, Manuel and Raffelt, Georg G.",
    title = "{Fast flavor pendulum: Instability condition}",
    eprint = "2412.09027",
    archivePrefix = "arXiv",
    primaryClass = "hep-ph",
    doi = "10.1103/PhysRevD.111.083028",
    journal = "Phys. Rev. D",
    volume = "111",
    number = "8",
    pages = "083028",
    year = "2025"
}

@article{Fiorillo:2025ank,
    author = "Fiorillo, Damiano F. G. and Raffelt, Georg G.",
    title = "{Theory of neutrino slow flavor evolution. Part II. Space-time evolution of linear instabilities}",
    eprint = "2501.16423",
    archivePrefix = "arXiv",
    primaryClass = "hep-ph",
    doi = "10.1007/JHEP06(2025)146",
    journal = "JHEP",
    volume = "06",
    pages = "146",
    year = "2025"
}

@article{Johns:2024bob,
    author = "Johns, Lucas",
    title = "{Implications of conservation laws and ergodicity for neutrino flavor instability}",
    eprint = "2402.08896",
    archivePrefix = "arXiv",
    primaryClass = "hep-ph",
    reportNumber = "LA-UR-24-21158",
    doi = "10.1103/qd8x-29zm",
    journal = "Phys. Rev. D",
    volume = "112",
    number = "6",
    pages = "063029",
    year = "2025"
}

@article{Cornelius:2023eop,
    author = "Cornelius, Marie and Shalgar, Shashank and Tamborra, Irene",
    title = "{Perturbing fast neutrino flavor conversion}",
    eprint = "2312.03839",
    archivePrefix = "arXiv",
    primaryClass = "astro-ph.HE",
    doi = "10.1088/1475-7516/2024/02/038",
    journal = "JCAP",
    volume = "02",
    pages = "038",
    year = "2024"
}

@article{Shalgar:2022lvv,
    author = "Shalgar, Shashank and Tamborra, Irene",
    title = "{Neutrino flavor conversion, advection, and collisions: Toward the full solution}",
    eprint = "2207.04058",
    archivePrefix = "arXiv",
    primaryClass = "astro-ph.HE",
    doi = "10.1103/PhysRevD.107.063025",
    journal = "Phys. Rev. D",
    volume = "107",
    number = "6",
    pages = "063025",
    year = "2023"
}

@article{Fiorillo:2024bzm,
    author = "Fiorillo, Damiano F. G. and Raffelt, Georg G.",
    title = "{Theory of neutrino fast flavor evolution. Part~I. Linear response theory and stability conditions.}",
    eprint = "2406.06708",
    archivePrefix = "arXiv",
    primaryClass = "hep-ph",
    doi = "10.1007/JHEP08(2024)225",
    journal = "JHEP",
    volume = "08",
    pages = "225",
    year = "2024"
}

@article{Bhattacharyya:2020jpj,
    author = "Bhattacharyya, Soumya and Dasgupta, Basudeb",
    title = "{Fast Flavor Depolarization of Supernova Neutrinos}",
    eprint = "2009.03337",
    archivePrefix = "arXiv",
    primaryClass = "hep-ph",
    reportNumber = "TIFR/TH/20-33",
    doi = "10.1103/PhysRevLett.126.061302",
    journal = "Phys. Rev. Lett.",
    volume = "126",
    number = "6",
    pages = "061302",
    year = "2021"
}

@article{Xiong:2023vcm,
    author = "Xiong, Zewei and Wu, Meng-Ru and Abbar, Sajad and Bhattacharyya, Soumya and George, Manu and Lin, Chun-Yu",
    title = "{Evaluating approximate asymptotic distributions for fast neutrino flavor conversions in a periodic 1D box}",
    eprint = "2307.11129",
    archivePrefix = "arXiv",
    primaryClass = "astro-ph.HE",
    doi = "10.1103/PhysRevD.108.063003",
    journal = "Phys. Rev. D",
    volume = "108",
    number = "6",
    pages = "063003",
    year = "2023"
}

@article{Duan:2006an,
    author = "Duan, Huaiyu and Fuller, George M. and Carlson, J and Qian, Yong-Zhong",
    title = "{Simulation of coherent nonlinear neutrino flavor transformation in the supernova environment: Correlated neutrino trajectories}",
    eprint = "astro-ph/0606616",
    archivePrefix = "arXiv",
    reportNumber = "LA-UR-06-4274",
    doi = "10.1103/PhysRevD.74.105014",
    journal = "Phys. Rev. D",
    volume = "74",
    pages = "105014",
    year = "2006"
}

@article{Xiong:2024pue,
    author = "Xiong, Zewei and Wu, Meng-Ru and George, Manu and Lin, Chun-Yu",
    title = "{Robust Integration of Fast Flavor Conversions in Classical Neutrino Transport}",
    eprint = "2403.17269",
    archivePrefix = "arXiv",
    primaryClass = "astro-ph.HE",
    doi = "10.1103/PhysRevLett.134.051003",
    journal = "Phys. Rev. Lett.",
    volume = "134",
    number = "5",
    pages = "051003",
    year = "2025"
}

@article{Tamborra:2020cul,
    author = "Tamborra, Irene and Shalgar, Shashank",
    title = "{New Developments in Flavor Evolution of a Dense Neutrino Gas}",
    eprint = "2011.01948",
    archivePrefix = "arXiv",
    primaryClass = "astro-ph.HE",
    doi = "10.1146/annurev-nucl-102920-050505",
    journal = "Annu. Rev. Nucl. Part. Sci.",
    volume = "71",
    pages = "165--188",
    year = "2021"
}

@article{Volpe:2023met,
    author = "Volpe, M. Cristina",
    title = "{Neutrinos from dense environments: Flavor mechanisms, theoretical approaches, observations, and new directions}",
    eprint = "2301.11814",
    archivePrefix = "arXiv",
    primaryClass = "hep-ph",
    doi = "10.1103/RevModPhys.96.025004",
    journal = "Rev. Mod. Phys.",
    volume = "96",
    number = "2",
    pages = "025004",
    year = "2024"
}

@article{Samuel:1993uw,
    author = "Samuel, Stuart",
    title = "{Neutrino oscillations in dense neutrino gases}",
    reportNumber = "IUHET-244",
    doi = "10.1103/PhysRevD.48.1462",
    journal = "Phys. Rev. D",
    volume = "48",
    pages = "1462--1477",
    year = "1993"
}

@article{Fiorillo:2024qbl,
    author = "Fiorillo, Damiano F. G. and Raffelt, Georg G.",
    title = "{Fast Flavor Conversions at the Edge of Instability in a Two-Beam Model}",
    eprint = "2403.12189",
    archivePrefix = "arXiv",
    primaryClass = "hep-ph",
    doi = "10.1103/PhysRevLett.133.221004",
    journal = "Phys. Rev. Lett.",
    volume = "133",
    number = "22",
    pages = "221004",
    year = "2024"
}

@article{Fiorillo:2024uki,
    author = "Fiorillo, Damiano F. G. and Raffelt, Georg G.",
    title = "{Theory of neutrino fast flavor evolution. Part II. Solutions at the edge of instability}",
    eprint = "2409.17232",
    archivePrefix = "arXiv",
    primaryClass = "hep-ph",
    doi = "10.1007/JHEP12(2024)205",
    journal = "JHEP",
    volume = "12",
    pages = "205",
    year = "2024"
}

@article{Zaizen:2022cik,
    author = "Zaizen, Masamichi and Nagakura, Hiroki",
    title = "{Simple method for determining asymptotic states of fast neutrino-flavor conversion}",
    eprint = "2211.09343",
    archivePrefix = "arXiv",
    primaryClass = "astro-ph.HE",
    doi = "10.1103/PhysRevD.107.103022",
    journal = "Phys. Rev. D",
    volume = "107",
    number = "10",
    pages = "103022",
    year = "2023"
}

@article{Zaizen:2023ihz,
    author = "Zaizen, Masamichi and Nagakura, Hiroki",
    title = "{Characterizing quasisteady states of fast neutrino-flavor conversion by stability and conservation laws}",
    eprint = "2304.05044",
    archivePrefix = "arXiv",
    primaryClass = "astro-ph.HE",
    doi = "10.1103/PhysRevD.107.123021",
    journal = "Phys. Rev. D",
    volume = "107",
    number = "12",
    pages = "123021",
    year = "2023"
}

@article{Fiorillo:2023hlk,
    author = "Fiorillo, Damiano F. G. and Raffelt, Georg G.",
    title = "{Flavor solitons in dense neutrino gases}",
    eprint = "2303.12143",
    archivePrefix = "arXiv",
    primaryClass = "hep-ph",
    doi = "10.1103/PhysRevD.107.123024",
    journal = "Phys. Rev. D",
    volume = "107",
    number = "12",
    pages = "123024",
    year = "2023"
}

@article{Pantaleone:1992eq,
    author = "Pantaleone, James T.",
    title = "{Neutrino oscillations at high densities}",
    reportNumber = "DOE-ER-40561-056, INT-92-07-01",
    doi = "10.1016/0370-2693(92)91887-F",
    journal = "Phys. Lett. B",
    volume = "287",
    pages = "128--132",
    year = "1992"
}

@article{Fiorillo:2023mze,
    author = "Fiorillo, Damiano F. G. and Raffelt, Georg G.",
    title = "{Slow and fast collective neutrino oscillations: Invariants and reciprocity}",
    eprint = "2301.09650",
    archivePrefix = "arXiv",
    primaryClass = "hep-ph",
    doi = "10.1103/PhysRevD.107.043024",
    journal = "Phys. Rev. D",
    volume = "107",
    number = "4",
    pages = "043024",
    year = "2023"
}

@article{Morinaga:2021vmc,
    author = "Morinaga, Taiki",
    title = "{Fast neutrino flavor instability and neutrino flavor lepton number crossings}",
    eprint = "2103.15267",
    archivePrefix = "arXiv",
    primaryClass = "hep-ph",
    doi = "10.1103/PhysRevD.105.L101301",
    journal = "Phys. Rev. D",
    volume = "105",
    number = "10",
    pages = "L101301",
    year = "2022"
}

@article{Dasgupta:2021gfs,
    author = "Dasgupta, Basudeb",
    title = "{Collective Neutrino Flavor Instability Requires a Crossing}",
    eprint = "2110.00192",
    archivePrefix = "arXiv",
    primaryClass = "hep-ph",
    reportNumber = "TIFR/TH/21-15",
    doi = "10.1103/PhysRevLett.128.081102",
    journal = "Phys. Rev. Lett.",
    volume = "128",
    number = "8",
    pages = "081102",
    year = "2022"
}

@article{Johns:2024esf,
    author = "Johns, Lucas",
    title = "{Implications of conservation laws and ergodicity for neutrino flavor instability}",
    eprint = "2402.08896",
    archivePrefix = "arXiv",
    primaryClass = "hep-ph",
    reportNumber = "LA-UR-24-21158",
    doi = "10.1103/qd8x-29zm",
    journal = "Phys. Rev. D",
    volume = "112",
    number = "6",
    pages = "063029",
    year = "2025"
}

@article{Fiorillo:2025gkw,
    author = "Fiorillo, Damiano F. G. and Janka, Hans-Thomas and Raffelt, Georg G.",
    title = "{Neutrino-Mass-Driven Instabilities as the Earliest Flavor Conversion in Supernovae}",
    eprint = "2507.22985",
    archivePrefix = "arXiv",
    primaryClass = "hep-ph",
    doi = "10.1103/jbmx-rbzt",
    journal = "Phys. Rev. Lett.",
    volume = "135",
    number = "23",
    pages = "231003",
    year = "2025"
}

@book{boyd2003physics,
  title={The Physics of Plasmas},
  author={Boyd, Thomas James Morrow and Sanderson, Jeffrey John},
  year={2003},
  publisher={Cambridge University Press}
}

@article{Bhattacharyya:2021klg,
    author = "Bhattacharyya, Soumya and Dasgupta, Basudeb",
    title = "{Fast flavor oscillations of astrophysical neutrinos with 1, 2, {\textellipsis}, {\ensuremath{\infty}} crossings}",
    eprint = "2101.01226",
    archivePrefix = "arXiv",
    primaryClass = "hep-ph",
    reportNumber = "TIFR/TH/20-47",
    doi = "10.1088/1475-7516/2021/07/023",
    journal = "JCAP",
    volume = "07",
    pages = "023",
    year = "2021"
}

@article{Raffelt:2025wty,
    author = "Raffelt, Georg G. and Janka, Hans-Thomas and Fiorillo, Damiano F. G.",
    title = "{Neutrinos from core-collapse supernovae}",
    eprint = "2509.16306",
    archivePrefix = "arXiv",
    primaryClass = "astro-ph.HE",
    month = "9",
    year = "2025",
    note = "To be published in the \textit{Encyclopedia of Particle Physics} (2026)"}

@article{Sarikas:2012ad,
    author = "Sarikas, Srdjan and de Sousa Seixas, David and Raffelt, Georg",
    title = "{Spurious instabilities in multi-angle simulations of collective flavor conversion}",
    eprint = "1210.4557",
    archivePrefix = "arXiv",
    primaryClass = "hep-ph",
    reportNumber = "MPP-2012-139",
    doi = "10.1103/PhysRevD.86.125020",
    journal = "Phys. Rev. D",
    volume = "86",
    pages = "125020",
    year = "2012"
}

@article{Samuel:1995ri,
    author = "Samuel, Stuart",
    title = "{Bimodal coherence in dense selfinteracting neutrino gases}",
    eprint = "hep-ph/9604341",
    archivePrefix = "arXiv",
    reportNumber = "MPI-PHT-95-57, CCNY-HEP-95-5",
    doi = "10.1103/PhysRevD.53.5382",
    journal = "Phys. Rev. D",
    volume = "53",
    pages = "5382--5393",
    year = "1996"
}

@article{Raffelt:2011yb,
    author = "Raffelt, Georg G.",
    title = "{N-mode coherence in collective neutrino oscillations}",
    eprint = "1103.2891",
    archivePrefix = "arXiv",
    primaryClass = "hep-ph",
    reportNumber = "MPP-2011-17, MPP-2011-17",
    doi = "10.1103/PhysRevD.83.105022",
    journal = "Phys. Rev. D",
    volume = "83",
    pages = "105022",
    year = "2011",
    note = "Erratum: \href{https://doi.org/10.1103/PhysRevD.104.089902}{{\em Phys. Rev. D} {\bf 104} (2021) 089902}"
}

@article{Duan:2005cp,
    author = "Duan, Huaiyu and Fuller, George M. and Qian, Yong-Zhong",
    title = "{Collective neutrino flavor transformation in supernovae}",
    eprint = "astro-ph/0511275",
    archivePrefix = "arXiv",
    doi = "10.1103/PhysRevD.74.123004",
    journal = "Phys. Rev. D",
    volume = "74",
    pages = "123004",
    year = "2006"
}

@article{Fiorillo:2025zio,
    author = "Fiorillo, Damiano F. G. and Raffelt, Georg G.",
    title = "{Dispersion relation of the neutrino plasma: Unifying fast, slow, and collisional instabilities}",
    eprint = "2505.20389",
    archivePrefix = "arXiv",
    primaryClass = "hep-ph",
    month = "5",
    year = "2025"
}

@article{Wu:2017drk,
    author = "Wu, Meng-Ru and Tamborra, Irene and Just, Oliver and Janka, Hans-Thomas",
    title = "{Imprints of neutrino-pair flavor conversions on nucleosynthesis in ejecta from neutron-star merger remnants}",
    eprint = "1711.00477",
    archivePrefix = "arXiv",
    primaryClass = "astro-ph.HE",
    doi = "10.1103/PhysRevD.96.123015",
    journal = "Phys. Rev. D",
    volume = "96",
    number = "12",
    pages = "123015",
    year = "2017"
}

@article{Dasgupta:2025quc,
    author = "Dasgupta, Basudeb and Mukherjee, Dwaipayan",
    title = "{Sufficient and Necessary Conditions for Collective Neutrino Instability: Fast, Slow, and Mixed}",
    eprint = "2505.03886",
    archivePrefix = "arXiv",
    primaryClass = "hep-ph",
    reportNumber = "TIFR/TH/25-12",
    month = "5",
    year = "2025"
}

@article{Nagakura:2022kic,
    author = "Nagakura, Hiroki and Zaizen, Masamichi",
    title = "{Time-Dependent and Quasisteady Features of Fast Neutrino-Flavor Conversion}",
    eprint = "2206.04097",
    archivePrefix = "arXiv",
    primaryClass = "astro-ph.HE",
    doi = "10.1103/PhysRevLett.129.261101",
    journal = "Phys. Rev. Lett.",
    volume = "129",
    number = "26",
    pages = "261101",
    year = "2022"
}

\onecolumngrid
\newpage

\begin{center}
\textbf{\large End Matter: Derivation of the Dispersion Relation}
\end{center}

\bigskip

\twocolumngrid

A neutrino plasma exhibits collective excitations due to the weak field each neutrino sources. We usually denote the component of the weak field that induces flavor conversions as $\psi$, which, in the usual two-flavor density matrix $\varrho$, is the lower left off-diagonal component, whereas the upper right component is $\psi^*$. Using one or the other as the mean field of flavor coherence causes certain sign differences between our results and those in the older literature, but no physical difference.

More specifically, we use the current $\psi^\mu$ associated with $\psi$ defined by
\begin{equation}
    \psi^\mu(\br)=\sum_{\bK}e^{i\bK\cdot \br}\psi^\mu_\bK,
\end{equation}
where
\begin{equation}
    \psi^\mu_\bK=2\sum_{\bp'}\left\langle a^\dagger_{e,\bp'-\frac{\bK}{2}}a_{\mu, \bp'+\frac{\bK}{2}}\right\rangle v^{\prime\mu},
\end{equation}
where $v^{\prime\mu}=(1,\bv')$ is the four-vector associated with the velocity $\bv'$, essentially a unit vector for ultrarelativistic neutrinos. The average $\langle\cdots\rangle$ is here performed over the quantum state of the neutrinos. The quanta of the field described by $\psi^\mu(\br)$ are termed flavomons. The field depends also on time, and we denote its Fourier transform with respect to both time and space as $\psi^\mu_{\Omega,\bK}$.

A nonvanishing $\psi^\mu(\br)$, i.e., neutrinos being in a superposition of flavor states, induces a weak field that tends to turn other neutrinos into a flavor eigenstate. In linear response theory, this is described by a flavor dielectric tensor, relating the medium response to the field $\psi^\mu(\br)$. This tensor, introduced in Refs.~\cite{Fiorillo:2024bzm,Fiorillo:2025npi}, is
\begin{equation}
    \varepsilon^{\mu\nu}(\Omega,\bK)=g^{\mu\nu}-\mu\int \frac{D(E,\bv) v^\mu v^\nu}{\omega-\bk\cdot \bv-\omega_E+i\epsilon}dE d^2\bv,
\end{equation}
where $g_{\mu\nu}$ is the metric and $\omega=\Omega+\mu D_0$, $\bk=\bK+\mu \bD_1$ as in the main text. Here $D_0=\int d^2\bv dE\,D(E,\bv)$ is the integrated DLN, and $\bD_1=\int d^2\bv dE\,D(E,\bv) \bv$ its flux. The second term describes the flavor field produced by the medium in response to $\psi^\mu(\br)$. Self-consistency therefore requires
\begin{equation}
    \varepsilon^{\mu\nu}(\Omega,\bK)\psi_{\nu,\Omega,\bK}=0.
\end{equation}
This is the dispersion relation, enforcing a specific relation between the eigenfrequency $\Omega_\bK$ for a given $\bK$ and the eigenvectors $e^\mu_{\bK}$ satisfying the consistency condition $\varepsilon^{\mu\nu}(\Omega_\bK,\bK)e_{\nu, \bK}=0$. 

This dispersion relation can be derived diagrammatically, by associating flavomons to quasi-particles in the medium coupled to the neutrinos with a strength $\sqrt{\sqrt{2}G_F |\mathcal{Z}_\bK|}\, (e_\bK\cdot v)$. Here $\mathcal{Z}_\bK$ is the wave-function renormalization of the flavomon, given explicitly in Ref.~\cite{Fiorillo:2025npi}.
 
The typical scale for the flavomon frequency and wave\-vector is defined by the total amount of DLN $\epsilon=D_0$.
In terms of this asymmetry parameter, we have $\omega, |\bk|\sim \mu \epsilon$.

The growth rate can then be computed as the difference between the rates for $\overline{\nu}_e\to \overline{\nu}_\mu+\psi$ and the inverse process. According to Ref.~\cite{Fiorillo:2025npi}, this leads to a growth rate of flavomons with shifted wave vector~$\bK$ of
\begin{equation}\label{eq:growth_rate_Feynman}
    \gamma_\bK=-\frac{\mathrm{Im}(\varepsilon_{\mu\nu})e^\mu_\bK e^{*\nu}_\bK}{\partial_\Omega \varepsilon_{\mu\nu}e^{\mu}_{\bK}e^{*\nu}_{\bK}}.
\end{equation}
This result descends purely from Fermi's Golden Rule for the processes of emission and absorption. It can also be derived from a direct expansion for small frequencies of the dispersion relation~\cite{Fiorillo:2024uki}, and we can now use it to obtain Eq.~\eqref{eq:approx_growth} reported in the main text. 

To this end, we restrict the system to axial symmetry, for which the dielectric tensor strongly simplifies to
\begin{widetext}
\begin{equation}
    \varepsilon^\mu_\nu=\begin{pmatrix}
        1- I_0 &  I_1 & 0 & 0\\
         I_1 & -1- I_2 & 0 & 0\\
        0 & 0 & -1-\frac{1}{2}(I_0-I_2) & 0\\
        0 & 0 & 0 &-1-\frac{1}{2}(I_0-I_2)
    \end{pmatrix}.
\end{equation}
\clearpage
\end{widetext}

For near-luminal flavomons, with $\omega\simeq k$, the integrals $I_n$, defined in Eq.~\eqref{eq:integrals}, are approximated by a relatively simple expression~\cite{Fiorillo:2024pns,Fiorillo:2025zio}
\begin{equation}
    I_n(\omega,k)\simeq \mu\left(\mathcal{I}(\omega,k)+\frac{d_n(k)}{k}\right),
\end{equation}
where
\begin{equation}
    \mathcal{I}(\omega,k)=\int dE\,\frac{D(E,1)}{k}\log\left(\frac{2k}{\omega-k-\omega_E}\right),
\end{equation}
and
\begin{equation}
    d_n(k)=\int dv dE\,\frac{D(E,v) v^n-D(E,1)}{1-v}. 
\end{equation}
The imaginary part of these integrals is even easier to understand from their original definition in Eq.~\eqref{eq:integrals}; expanding the $i\epsilon$ with the delta function, we see that
\begin{equation}\label{eq:imaginary_part}
    \mathrm{Im}(I_n)=\pi \mu\int dv dE\,D(E,v) v^n \delta(\omega-kv-\omega_E).
\end{equation}
In the Feynman diagram language, the delta function enforces energy conservation, as required by Fermi's Golden Rule. For $\omega\sim k$, since $\omega_E\ll \omega, k$ (we are focusing only on gapped modes with $\omega, k$ given by the refractive energy scale, which is much larger than the vacuum frequency), $v$ is very close to $1$, so we may replace $v^n\simeq 1$. The imaginary part is therefore the same for all values of $n$ to a first approximation so that $\mathrm{Im}(I_n)\simeq \mathrm{Im}(\mathcal{I})$.

We focus on the longitudinal modes (L1 and L2 in the main text) which are polarized only longitudinally to the flavomon, excluding the transverse modes T that have $e_\bK=(0,0,1,0)$ or $(0,0,0,1)$, assuming propagation in the $x$ direction. For the longitudinal modes, we have
\begin{equation}
    e^0_\bK=1
    \quad\textrm{and}\quad
    e^1_{\bK}=\frac{\mathrm{Re}(I_0)-1}{\mathrm{Re}(I_1)}.
\end{equation}
(We have to assume as usual $\gamma_\bK\ll \omega-k$ so that flavomons are well defined.) We can now explicitly replace these expressions in Eq.~\eqref{eq:growth_rate_Feynman} to obtain
\begin{equation}
    \gamma_\bK=-\frac{\mathrm{Im}(\mathcal{I})}{\partial_\Omega \mathcal{I}}.
\end{equation}
In the denominator, performing the derivative, gives
\begin{equation}
    \partial_\Omega\mathcal{I}=-\fint dE\,\frac{D(E,1)}{k(\omega-k-\omega_E)}.
\end{equation}
Instead, the numerator, after explicitly integrating the delta function in Eq.~\eqref{eq:imaginary_part}, gives
\begin{equation}\label{eq:Im_I}
    \mathrm{Im}(\mathcal{I})=\pi\int dE\,\frac{D(E,1)}{k}\,
    \Theta\left(1-\frac{\mathrm{Re}(\omega)-\omega_E}{k}\right).
\end{equation}
This expression finally coincides with Eq.~\eqref{eq:approx_growth} reported in the main text. This derivation, however, shows how physical insight translates into mathematics: by integrating over the delta function of energy conservation in the flavomon production, only the neutrinos kinematically allowed to decay contribute to the growth, explaining the Heaviside function in Eq.~\eqref{eq:Im_I}. 

The signs of $\partial_\Omega \mathcal{I}$ and $\mathrm{Im}(\mathcal{I})$ are difficult to infer from the mathematical expressions, but if the amount of flipped neutrinos added to the plasma is sufficiently small, the signs can be understood from a continuity argument. For a plasma containing only $\nu_e$ (or potentially $\overline{\nu}_\mu$), there can be no instability. In this case, $D(E,v)>0$, so $\mathrm{Im}(\mathcal{I})$ has the same sign as $k$; in order not to have any instability, also $\partial_\Omega \mathcal{I}$ has the same sign as $k$. When a small amount of flipped neutrinos are added to the plasma, $\partial_\Omega\mathcal{I}$ suffers only a small change, since it depends on a global integral of the DLN distribution, so an instability must be associated with values of $\omega$ and $k$ for which $\mathrm{Im}(\mathcal{I})k<0$.

This derivation also reveals further insight into the generality of our conclusion. An unstable mode, if appearing anywhere, must first transition through a region where $\mathrm{Im}(\omega)\simeq 0$. In this region, the approximation of treating it as a flavomon with a well-defined energy, i.e., Eq.~\eqref{eq:growth_rate_Feynman}, is exact. Then we only need to understand under what condition this equation can predict a growth rate that is very small, yet positive. If both $|\omega|, |k| \gg |\omega_E|$ holds (the gapped modes we are considering), then the only way to satisfy energy conservation
\textit{and} to have a small but nonvanishing growth rate, is $\omega\simeq k$. Near-luminal gapped flavomons are therefore the only ones that can, in principle, become unstable. For them, our argument based on Eq.~\eqref{eq:approx_growth} is exact, as confirmed by the perfect agreement with the analytical approximation shown in~Fig.~\ref{fig:analytic}. 

In principle, modes with $\omega<k$ (subluminal modes) can also satisfy energy conservation; but since for $\omega_E=0$ these modes are always strongly Landau damped with a rate $\gamma_\bK\sim \mu \epsilon$ in the absence of angular crossing~\cite{Fiorillo:2024bzm,Fiorillo:2024uki, Fiorillo:2025zio}, they cannot exhibit an instability for $\omega_E\ll \mu \epsilon$. 

There is a final possibility, namely that $\omega, k\sim \omega_E$, so that energy conservation can be enforced for any value of $\omega$ and $k$ for some energy $E$. Such gapless modes are very difficult to study, but this is generally unnecessary. Gapless modes exist only in a very narrow range of frequency and wave vectors, so they correspond to waves with very large wavelength, reaching up to kilometers. In a real environment, the matter and neutrino density is inhomogeneous over these distances, so these modes, strictly speaking, do not exist. Therefore, we do not devote further attention to determining their existence or potential instabilities. 

\end{document}